# Bayesian computation via empirical likelihood

Kerrie L. Mengersen[1], Pierre Pudlo[2,3,4], and Christian P. Robert[5,6,7]

[1]Department of Statistics, QUT, Australia, [2]CBGP, INRA, Montpellier, [3]Université Montpellier 2, I3M, [4]Institut de Biologie Computationnelle, Montpellier, France, [5]Université Paris Dauphine, CEREMADE, [6]Institut Universitaire de France, [7]CREST, Paris, France

October 30, 2018

**Abstract**

Approximate Bayesian computation (ABC) has become an essential tool for the analysis of complex stochastic models when the likelihood function is numerically unavailable. However, the well-established statistical method of empirical likelihood provides another route to such settings that bypasses simulations from the model and the choices of the ABC parameters (summary statistics, distance, tolerance), while being convergent in the number of observations. Furthermore, bypassing model simulations may lead to significant time savings in complex models, for instance those found in population genetics. The $BC_{el}$ algorithm we develop in this paper also provides an evaluation of its own performance through an associated effective sample size. The method is illustrated using several examples, including estimation of standard distributions, time series, and population genetics models.

## 1 Introduction

Bayesian statistical inference cannot easily operate when the likelihood function associated with the data is not entirely known, or cannot be computed in a manageable time, as is the case in most population genetic models (1, 2, 3). The fundamental reason for this difficulty with population genetics is that the statistical model associated with coalescent data needs to integrate over trees of high complexity. Similar computational issues with the likelihood function often occur in hidden Markov and other dynamic models (4). In those settings, traditional approximation tools based on stochastic simulation (5) are unavailable or unreliable. Indeed, the complexity of the latent structure defining the likelihood makes simulation of such structures too unstable to be trusted. Such settings call for alternative and often cruder approximations. The ABC methodology (1, 6) is a popular solution that bypasses the computation of the likelihood function (see (7) and (8) for surveys); (9) validate a conditional version of ABC that applies to hierarchical Bayes models in a wide generality.

The fast and polytomous development of the ABC algorithm is indicated by the rising literature in the domain, at both the methodological and the application levels. For instance, a whole new area of population genetic modelling (10, 8) has been explored thanks to the availability of such methods. However, both practitioners and theoreticians show a reluctance in adopting ABC, as some doubt about the validaty of the method (11, 12, 13). We propose in this paper to supplement the ABC approach with a generic and convergent likelihood approximation called the empirical likelihood that validates the new Bayesian computational technique as a convergent inferential method when the number of observations grows to infinity. The empirical likelihood perspective, introduced by (14), is a robust statistical approach that does not require the specification of the likelihood function. However, while it does not appear to have been used before in the settings that now rely on ABC, this data analysis method also is a broadly (albeit not universally) applicable and often fast approach which approximation differs from the one found in ABC algorithms, even though both are rooted in non-parametric statistics. Therefore, this methodology can be used both as a solution *per se* and as a benchmark against which to test the ABC output in many cases. This paper introduces the BC$_{el}$ algorithm and illustrates its performances on selected representative examples, comparing the outcome with the true posterior density whenever available, and with an ABC approximation (15) otherwise.

## 2 Statistical Methods

### 2.1 The ABC algorithm

The primary purpose of the ABC algorithm is to approximate simulation from the centerpiece of Bayesian inference, the posterior distribution $\pi(\boldsymbol{\theta}|\mathbf{y}) \propto \pi(\boldsymbol{\theta})f(\mathbf{y}|\boldsymbol{\theta})$, when it cannot be numerically computed but when the distributions corresponding to both the prior $\pi$ and the likelihood $f$ can be simulated by manageable computer devices. The original (6) ABC algorithm is as follows: given a sample $\mathbf{y}$ of observations from the sample space, a sample of parameters $(\boldsymbol{\theta}_1, \ldots, \boldsymbol{\theta}_M)$ is produced by

**Algorithm 1: ABC sampler**
  for $i = 1$ to $M$ do
    repeat
      Generate $\boldsymbol{\theta}'$ from the prior distribution $\pi(\cdot)$
      Generate $\mathbf{z}$ from the likelihood $f(\cdot|\boldsymbol{\theta}')$
    until $\rho\{\eta(\mathbf{z}), \eta(\mathbf{y})\} \leq \epsilon$
    set $\boldsymbol{\theta}_i = \boldsymbol{\theta}'$,
  end for

The parameters of the ABC algorithm are the summary statistic $\eta$, the distance $\rho\{\cdot, \cdot\}$ and the tolerance level $\epsilon > 0$. The basic justification of the ABC approximation is that, when using a sufficient statistic $\eta$, the distribution of the $\boldsymbol{\theta}_i$'s in the output of the algorithm converges to the genuine posterior distribution when $\epsilon$ goes to zero (16).

In practice, however, the statistic $\eta$ is non-sufficient and at best the approximation then converges to the genuine posterior $\pi(\boldsymbol{\theta}|\eta(\mathbf{y}))$ when $\epsilon$ goes to zero. This loss of information seems to be a necessary price to pay for the access to computable quantities. Furthermore, as argued below, it can be evaluated against the empirical likelihood approximation when the latter is available. Indeed, this approach does not require an information reduction through the choice of a tolerance zone or of a non-sufficient summary statistic.

### 2.2 Empirical likelihood

Owen (14) developed empirical likelihood techniques as a robust alternative to classical likelihood approaches. He demonstrated that, for some categories of statistical models, this approach inherited the convergence properties of standard likelihood at a much lower cost in assumptions about the model (as detailed in SI). While ABC algorithms do require a fully defined and often complex (hence debatable) statistical model, we argue that one should take advantage of the approximation device provided by the empirical likelihood to overcome most of the calibration difficulties encountered by ABC, at least as a convenient benchmark against which to test ABC solutions.

Assume that the dataset $\mathbf{y}$ is composed of $n$ independent replicates $\mathbf{y} = (\mathbf{y}_1, \ldots, \mathbf{y}_n)$ of some random vector $Y$ with density $f$. Rather than defining the likelihood from the density $f$ as usual, the empirical likelihood method starts by defining parameters of interest, $\boldsymbol{\theta}$, as functionals of $f$, for instance as moments of $f$, and it then profiles a non-parametric likelihood. More precisely, given a set of constraints of the form

$$\mathbb{E}_F[h(Y, \theta)] = 0, \tag{1}$$

where the dimension of $h$ sets the number of constraints unequivocally defining $\boldsymbol{\theta}$, the empirical likelihood is defined as

$$L_{el}(\boldsymbol{\theta}|\mathbf{y}) = \max_{\mathbf{p}} \prod_{i=1}^{n} p_i \tag{2}$$

for $\mathbf{p}$ in the set $\{\mathbf{p} \in [0; 1]^n, \sum p_i = 1, \sum_i p_i h(\mathbf{y}_i, \boldsymbol{\theta}) = 0\}$. For instance, in the one-dimensional case when $\theta = \mathbb{E}_f[Y]$, the empirical likelihood in $\theta$ is the maximum of the product $p_1 \cdots p_n$ under the constraint $p_1 y_1 + \ldots + p_n y_n = \theta$. (Solving (2) is done using the R package 'emplik' developed by (17) and based on the Newton-Lagrange algorithm.) When the data are not iid, an underlying iid structure may sometimes be exploited, as illustrated in the dynamic model section below. However, this is not always the case, meaning that the empirical likelihood method remains out of reach in some complex cases when ABC can still be implemented. Furthermore, as pointed out in the SI by a quote from Owen, the validation of the approach depends on a choice of the set of constraints that ensures convergence.

While the convergence of the empirical likelihood is well-established (see SI and (18)), the Bayesian use of empirical likelihoods has been little examined in the past, apart from a Monte Carlo study in (19), and a probabilistic one in (20).

## 2.3 BC$_{el}$

The most natural use of the empirical likelihood approximation is to act as if this representation was an exact likelihood, as in (19). Incorporating this perspective into a basic sampler leads to the following algorithm:

**Algorithm 2: Basic BC$_{el}$ sampler**

  **for** $i = 1$ to $M$ **do**
    Generate $\boldsymbol{\theta}_i$ from the prior distribution $\pi(\cdot)$
    Set the weight $\omega_i = L_{el}(\boldsymbol{\theta}_i|\mathbf{y})$
  **end for**

The output of BC$_{el}$ is a sample of size $M$ of parameters with associated weights, which operate as an importance sampling output (5). Thus, the performance of the algorithm can be evaluated through the effective sample size

$$\text{ESS} = 1 \bigg/ \sum_{i=1}^{M} \left\{ \omega_i \bigg/ \sum_{j=1}^{M} \omega_j \right\}^2,$$

which approximates the size of an iid sample with the same variance as the original sample. As shown in (21), this quantity is always between 1 (corresponding to a very poor outcome) and $M$ (corresponding to an iid perfect outcome).

Any algorithm that samples from a posterior distribution (e.g., MCMC, Population Monte Carlo, SMC algorithms, see (5)) may instead use the empirical likelihood as a proxy to the exact likelihood. For instance, to speed up the computation in the population genetics model introduced below, we resorted to the adaptive multiple importance sampling (AMIS, (22)) which is easy to parallelize on a multi-core computer. While the original target distribution is $\pi(\boldsymbol{\theta})L(\boldsymbol{\theta}|\mathbf{y})$ and the AMIS algorithm uses several (multivariate) Student's $t$ distributions, denoted $t_3(\cdot|\mathbf{m}, \boldsymbol{\Sigma})$ (i.e., with three degrees of freedom, centered at mean $\mathbf{m}$ and with covariance matrix $\boldsymbol{\Sigma}$), as an importance sampling distribution, the algorithm can be adapted to the empirical likelihood in a straightforward manner:

**Algorithm 3: $BC_{el}$-AMIS sampler**

  **for** $i = 1$ to $M$ **do**
    Generate $\boldsymbol{\theta}_{1,i}$ from the prior distribution $q_1(\cdot)$
    Set $\omega_{1,i} = L_{el}(\boldsymbol{\theta}_{1,i}|\mathbf{y})$
  **end for**
  **for** $t = 2$ to $T_M$ **do**
    Compute weighted mean $\mathbf{m}_t$ and weighted variance matrix $\boldsymbol{\Sigma}_t$ of the $\boldsymbol{\theta}_{s,i}$ ($1 \leq s \leq t-1$, $1 \leq i \leq M$).
    Denote $q_t(\cdot)$ the density of $t_3(\cdot|\mathbf{m}_t, \boldsymbol{\Sigma}_t)$.
    **for** $i = 1$ to $M$ **do**
      Generate $\boldsymbol{\theta}_{t,i}$ from $q_t(\cdot)$.
      Set $\omega_{t,i} = \pi(\boldsymbol{\theta}_{t,i})L_{el}(\boldsymbol{\theta}_{t,i}|\mathbf{y}) / \sum_{s=1}^{t-1} q_s(\boldsymbol{\theta}_{t,i})$
    **end for**
    **for** $r = 1$ to $t - 1$ **do**
      **for** $i = 1$ to $M$ **do**
        Update the weight of $\boldsymbol{\theta}_{r,i}$ as
        $\omega_{r,i} = \pi(\boldsymbol{\theta}_{t,i})L_{el}(\boldsymbol{\theta}_{r,i}|\mathbf{y}) / \sum_{s=1}^{t-1} q_s(\boldsymbol{\theta}_{r,i})$
      **end for**
    **end for**
  **end for**

The output is thus a weighted sample $\boldsymbol{\theta}_{t,i}$ of size $MT_M$.

In contrast with ABC, $BC_{el}$ algorithms do not usually require simulations from the sampling model, given that (2) provides a converging and non-parametric approximation of the likelihood function. This feature thus induces very significative improvements in computing time when the production of pseudo-datasets is time consuming, since solving (1) is usually immediate. This is for instance the case in population genetics and the last section of the SI provides an illustration of a huge improvement in comparison with ABC in two experiments described below. However, the improvement in speed may vanish in cases when producing an iid structure connected with the constraint (1) requires simulations from the sampling model, as illustrated by a counter-example for point processes in the SI, $BC_{el}$ and ABC then breaking even in terms of computing time. Even though the computing time required by $BC_{el}$ is customarily negligible when compared with ABC (or does not induce any extra time as in the point process counter-example), we further caution against opposing both approaches solely based on computing times, since they differ in the approximations they provide to a genuine Bayesian analysis and thus should be used in conjunction.

Using empirical likelihoods means there is no calibration of the many tuning parameters of ABC algorithms; most significantly, the likelihood ratio acts as a natural distance and importance weights produce an implicit and self-defined quantile on the original sample simulated from the prior. Notwithstanding these appealing qualities, $BC_{el}$ still requires calibration, in particular in the choices of the parameterization of the sampling distribution and of the corresponding constraints (1) defining the empirical likelihood. Some examples are discussed below. The $BC_{el}$-AMIS sampler also implies computing values of the prior density, up to a constant, which may be an hindrance in peculiar cases.

## 2.4 Composite likelihood in population genetics

ABC was first introduced by population geneticists (2, 10, 6) interested in statistical inference about the evolutionary history of species, as no likelihood-based approach existed apart from very rudimentary and hence unrealistic situations. This approach has been used in a number of biological studies (23, 24, 25), most of them including model choice. It is therefore crucial to obtain insights into the validity of such studies, particularly when they have economic, biological or ecological consequences (see, e.g., (26)). This can be achieved in part by running a comparison using $BC_{el}$. Furthermore, given the major gain in computing time, due to the absence of replications of the data, $BC_{el}$ can be applied to more complex biological models.

The main caveat when using the empirical likelihood in such settings is selecting a constraint (1) on the parameter of interest: in phylogeography, parameters like divergence dates, effective population sizes, mutation rates, etc., cannot be expressed as moments of the sampling distribution at a given locus. In particular, the data are not iid. However, when considering microsatellite loci with the stepwise mutation model (27) and evolutionary scenarios composed of divergence, we can derive the pairwise composite scores whose zero is the pairwise maximum likelihood estimator. Composite likelihoods have been proved consistent for estimating recombination rates, introducing an approximation of the dependency structure between nearby loci (28, 29, 30, 31). (See also (32) for composite likelihoods used in a likelihood-free setting.)

More specifically, we are approximating the intra-locus likelihood by a product over all pairs of genes in the sample at a given locus. Assuming that $y_i^k$ denotes the allele of the $i$-th gene in the sample at the $k$-th locus, and that $\phi$ is the vector of parameters, then the so-called pairwise likelihood of the data at the $k$-th locus, namely $\mathbf{y}^k$, is defined by

$$\ell_2(\mathbf{y}^k|\phi) = \prod_{i<j} \ell_2(y_i^k, y_j^k|\phi)$$

and the corresponding pairwise score function is $\nabla_\phi \log \ell_2(\mathbf{y}^k|\phi)$. Pairwise score equations

$$\mathbb{E}_f[\nabla_\phi \log \ell_2(Y|\phi)] = 0$$

provide a constraint (1) in every way comparable to the score equations that give the maximum likelihood estimate and which is quite powerful for empirical likelihood derivations ((18), pp. 48–50). Hence the empirical likelihood of the full dataset $\mathbf{y} = (\mathbf{y}^1, \ldots, \mathbf{y}^K)$ given $\phi$ is computed with (2) under the (multidimensional) constraint that

$$\sum_{k=1}^{K} p_k \nabla_\phi \log \ell_2(\mathbf{y}^k|\phi) = 0.$$

When the effective population size is identical over all populations of the demographic scenario, the time axis may be scaled so that coalescence of two genes in Kingman's genealogy occurs with rate $k(k-1)/2$ if there are $k$ lineages. In this modified scale, mutations at a given locus arise with rate $\theta/2$ along the gene genealogy. Our mutation model is the simple stepwise mutation model of (27), i.e. the number of repeats of the mutated gene increases or decreases by one unit with equal probability. Given two microsatellite allelic states $x_1$ and $x_2$, their pairwise likelihood $\ell_2(x_1, x_2|\phi)$ depends only on the difference of the states $x_1 - x_2$. If both genes belong to individuals that lie in the same deme, then (see SI and (33))

$$\ell_2(x_1, x_2|\phi) = \rho(\theta)^{|x_2 - x_1|}/\sqrt{1 + 2\theta}$$

where $\rho(\theta) = \theta/(1 + \theta + \sqrt{1 + 2\theta})$. If the two genes belong to individuals from demes having diverged at time $\tau$, then (33)

$$\ell_2(x_1, x_2|\phi) = \frac{e^{-\tau\theta}}{\sqrt{1 + 2\theta}} \sum_{k=-\infty}^{+\infty} \rho(\theta)^{|k|} I_{|x_1 - x_2| - k}(\tau\theta)$$

where $I_\delta(z)$ denotes the $\delta$th-order modified Bessel function of the first kind evaluated at $z$. Computing the pairwise scores, i.e. partial derivatives of $\log \ell_2(x_1, x_2|\phi)$ from those equations, is straightforward, by virtue of recurrence properties of the Bessel functions. Algorithm $BC_{el}$ is therefore directly available in this setting, and furthermore at a cost much lower than the one associated with ABC algorithms (Table S1).

## 3 Results

### 3.1 Normal distribution

Starting with the benchmark of a normal distribution with known variance (equal to one), we can check that the empirical likelihood allows for a proper recovery of the true posterior distribution on the mean. Fig. S1 shows that a constraint (1) based on the mean works well, as do the two constraints on mean and second central moment, $\mathbb{E}[(X - \theta)^2] = 0$ (Figure S2). On the other hand, using the three first central moments in the empirical likelihood may degrade the fit (three cases in Fig. S3). While this poor fit is not signaled by the ESS (which is often larger than in Fig. S1–S2, because of the growing disconnection between the approximation and the true likelihood and hence a more uniform range of the weights), a parallel run of the method with different collections of constraints does detect the discrepancy. This illustrates the variability of the empirical likelihood approximation, as well as its sensitivity to the choice of defining constraints. While a drawback of the method, this variability can be tested and evaluated by comparing outcomes, due to often limited computing costs. This toy experiment also supports the generic recommendation (18) to keep the number of constraints and parameters equal.

### 3.2 Quantile distributions

Quantile distributions are defined by a closed-form quantile function $F^{-1}(p; \theta)$, and generally have no closed form for the density function. They are of great interest because of their flexibility and the ease with which they can be simulated by a simple inversion of the uniform distribution. A range of methods, including ABC approaches (10), have been proposed for estimation (see SI). We focus here on the four-parameter $g$-and-$k$ distribution, defined by its quantile function, denoted $Q(r; A, B, g, k)$ and equal to

$$A + B\left(1 + c\frac{1 - \exp(-gz(r))}{1 + \exp(-gz(r))}\right)\left(1 + z(r)^2\right)^k z(r)$$

where $z(r)$ is the $r$th standard normal quantile; the parameters $A, B, g$ and $k$ represent location, scale, skewness and kurtosis, respectively and $c$ measures the overall asymmetry (34, 35). We evaluated the $\text{BC}_{\text{el}}$ algorithm for estimating this distribution using two values of $\theta = (A, B, g, k)$, two sets of priors and various combinations of $n, M$ and $p$, where $p$ is the number of percentiles used as constraints (see details in SI).

Figure 1 illustrates the true and fitted curves and a 95% credible region for the case with $n = 100, M = 5000$ and $p = 3$. The corresponding posterior means (standard deviations) for the parameters $A, B, g, k$ were $3.08(0.14), 1.12(0.23), 1.79(0.25), 0.41(0.12)$, respectively. The choice of sample size and number of constraints did not substantively affect the accuracy of parameter estimates, but the precision was noticeably improved for the larger sample size; see Figures S4, S5, and S6.

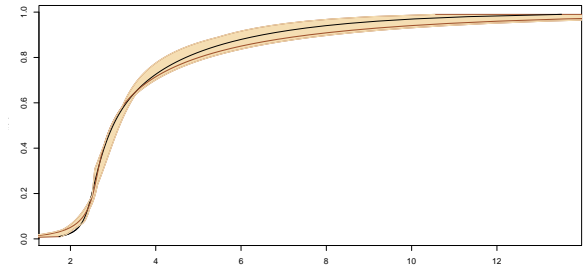

Figure 1: True (black) and fitted (brown) cdf functions with a pointwise 95% credible (shaded grey) region centered on the fitted cdf for a dataset of $n = 100$ observations from the $g$-and-$k$ distribution, based on $M = 10^3$ simulations of $\text{BC}_{\text{el}}$.

The accuracy and precision of the estimates were broadly comparable with the results obtained by (36) for the same distribution. Based on the whole experiment, the parameters $A$ and $B$ were well estimated in all cases, while the estimates of $g$ and $k$ were poorer for smaller values of $n$ and $M$. For small $n$ the estimates were more subject to the vagaries of sampling variation, whereas for small $M$ they were subject to the influence of a smaller number of very large importance weights. However, given the speed of $\text{BC}_{\text{el}}$ compared with competing ABC algorithms, it is feasible to use even larger values of $M$ than considered in this experiment, since there is no requirement to simulate new datasets at each iteration. Moreover, this experiment is based on the very basic case of sampling from the prior; the results would be further improved by using an analogue of $\text{BC}_{\text{el}}$-AMIS or alternative approaches similar to those proposed by (37) for ABC.

## 3.3 Dynamic models

In dynamic models, the difficulty with empirical likelihood stems from the dependence in the data $(y_t)_{1 \leq t \leq T}$. However, these models can be represented as transforms of unobserved iid sequences $(\epsilon_t)_{1 \leq t \leq T}$. The recovery of a converging empirical likelihood representation thus requires the reconstitution of the $\epsilon_t$'s as transforms of the data $\mathbf{y}$ and of the parameter $\theta$. Independence between the $\epsilon_t$'s is then at least as important as moment conditions. (This implies equivalent computing times for ABC and $\text{BC}_{\text{el}}$.)

For instance, consider a simple dynamic model, namely the ARCH(1) model:

$$y_t = \sigma_t \epsilon_t, \quad \epsilon_t \sim \mathcal{N}(0,1), \quad \sigma_t^2 = \alpha_0 + \alpha_1 y_{t-1}^2,$$

with a uniform prior over the simplex, i.e., $\alpha_0, \alpha_1 \geq 0, \alpha_0 + \alpha_1 \leq 1$. While this model can be handled by other means, since the likelihood function is available, we will compare here the behaviour of ABC and $\text{BC}_{\text{el}}$ algorithms.

First, a natural empirical likelihood representation is based on the reconstituted $\epsilon_t$'s, defined as $y_t/\sigma_t$ when the $\sigma_t$'s are derived recursively. Figure 2 shows the result of estimating both parameters $\alpha_0$ and $\alpha_1$ when Algorithm ABC uses as summary statistics either the least square estimates of the parameters (derived from the series $(y_t^2)$), which we label "optimal ABC" in connection with (38), or the mean of the series $\log(y_t^2)$ supplemented by the two first autocorrelations of the series $(y_t^2)$. The constraints in the empirical likelihood are either based on the three first moments of the reconstituted $\epsilon_t$'s or on the variance of those $\epsilon_t$'s complemented by both the correlations between the $y_{t-1}$'s and the $\epsilon_t$'s and between the $\epsilon_{t-1}$'s and the $\epsilon_t$'s. As seen from this experiment, $BC_{el}$ does as well as the optimal ABC for the estimation of the parameters, but further brings a reduction in the variability of those estimates, thanks to the importance weights.

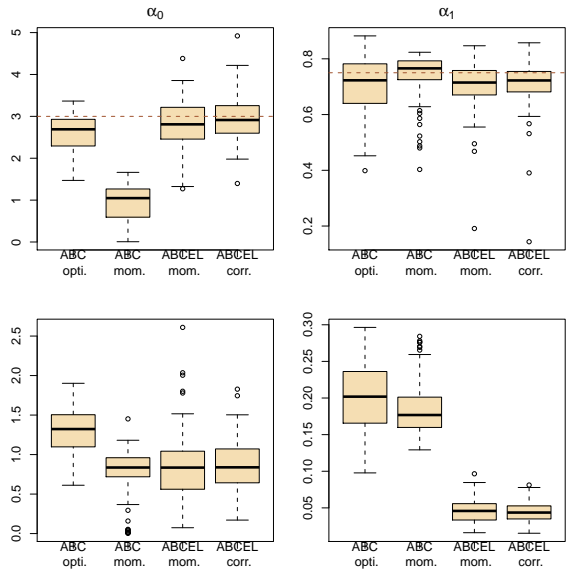

Figure 2: Comparison of ABC evaluations of posterior expectations *(top, with true values in dashed lines)* and posterior variances *(bottom)* of the parameters $(\alpha_0, \alpha_1)$ of the ARCH(1) model with 100 observations. The first two columns correspond to two choices of summary statistics for the ABC algorithm (least squares estimates and mean of the $\log y_t$'s plus autocorrelations of order 2 and 3, respectively). The last two columns correspond to two sets of constraints for the $BC_{el}$ alternative (first three moments and second moment plus autocorrelation of order 1 plus correlation with previous observation for the reconstituted $\epsilon_t$'s). All experiments are based on the same reference table of $10^4$ simulations, with the tolerance $\epsilon$ chosen as the 1% quantile of the distances.

A much more complex dynamic model is the GARCH(1,1) model of (39) that can be formalized as the observation of $y_t \sim \mathcal{N}(0, \sigma_t^2)$ when

$$\sigma_t^2 = \alpha_0 + \alpha_1 y_{t-1}^2 + \beta_1 \sigma_{t-1}^2 \qquad (3)$$

under the constraints $\alpha_0, \alpha_1, \beta_1 > 0$ and $\alpha_1 + \beta_1 < 1$, that is, $y_t = \sigma_t \epsilon_t$. Given the constraints on the parameters, a natural prior is to choose an exponential distribution on $\alpha_0$, for instance an exponential $\mathcal{E}xp(1)$ distribution, and a Dirichlet $\mathcal{D}_3(1,1,1)$ on $(\alpha_1, \beta_1, 1 - \alpha_1 - \beta_1)$. An ABC approach requires the choice of summary statistics, which are necessarily non-sufficient since the model is a state-space model. Following (38), we use the maximum likelihood estimator as summary statistics, relying on the R function `garch` for its derivation despite its lack of stability. Since (40) derived natural score constraints for the empirical likelihood associated with this model, we used their constraints to build our BC$_{\text{el}}$ algorithm. Fig. 3 provides a comparison of both approaches with the MLE. It shows in particular that the ABC algorithm is unable to produce acceptable inference in this case, even in the most favorable case when it is initialized at a satisfactory maximum likelihood estimate (as shown by the bottom row). The BC$_{\text{el}}$ algorithm is performing better, even though it fails to catch the correct range of $\beta_1$.

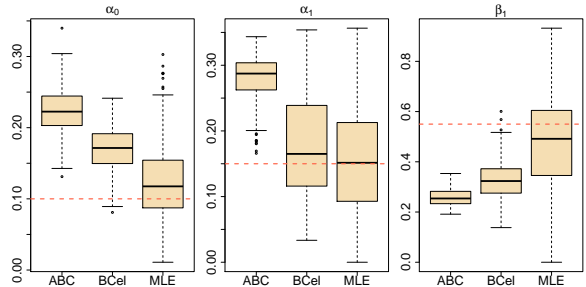

Figure 3: Comparison of evaluations of posterior expectations *(with true values in dashed lines)* of the parameters $(\alpha_0, \alpha_1, \beta_1)$ of the GARCH(1) model with 250 observations. The first row corresponds to an optimal ABC algorithm (using the MLE as summary statistic and with the tolerance $\epsilon$ chosen as the 5% quantile of the distances), the second row corresponds to the BC$_{\text{el}}$ algorithm based on the constraints derived in (40), and the third row corresponds to the MLE derived by the R procedure garch initialized at the true parameter value.

Another type of non-iid model relying on the superposition of an unknown number of gamma point processes and processed in (41) through a (non-Bayesian) alternative to ABC is discussed in the SI as an additional illustration of the possibilities of the empirical likelihood perspective for complex models, offering a free benchmark for evaluating the ABC outcome. Figure S7 shows a clear improvement brought by BC$_{\text{el}}$ over the corresponding ABC outcome.

### 3.4 Population genetics

We compare our proposal with the reliable ABC-based estimates given by (3). We set up two toy experiments that are designed to defeat ABC, using pseudo observed data. The two evolutionary scenarios are given in Figure 4. In all experiments, we only consider microsatellite loci and assume that the effective population size is identical over all populations of the scenario.

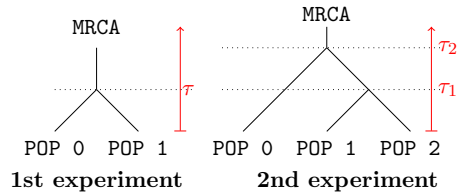

Figure 4: Evolutionary scenarios of the two experiments in population genetics.

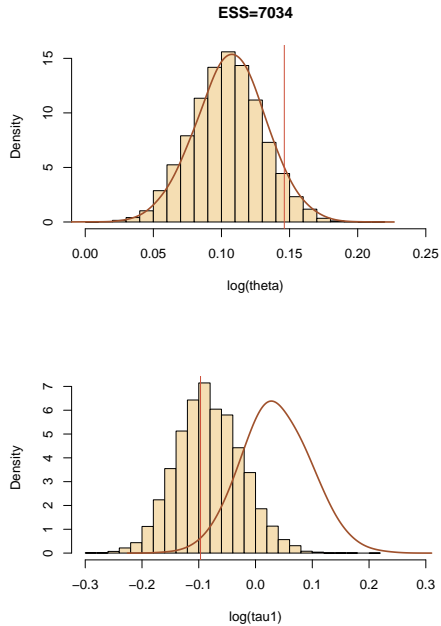

Figure 5: Comparison of the original ABC (curve) with the histogram of the simulation from the $BC_{el}$–AMIS sampler in the case of the population genetics model given in Scenario A, based on uniform priors on $(\log_{10}(\theta), \log_{10}(\tau))$ on $(-1, 1.5) \times (-1, 1)$ and $10^4$ particles.

In the first experiment, we consider two populations which diverged at time $\tau$ in the past, see Figure 4 *(left)*. Our pseudo observed datasets are made of thirty diploid individuals per population genotyped at a hundred independent loci. We compare the marginal posterior distributions of the unknown parameters $\theta$ and $\tau$ computed with the ABC method (using the DIY-ABC software of (42)) and with the $BC_{el}$-AMIS sampler. In this case, results are improved when the $\theta$-component of the composite scores, namely $\partial_\theta \log \ell_2(\mathcal{D}|\phi)$, is restricted to the sum over all pairs of genes lying in *the same population*. Otherwise, as can be checked via a quick simulation experiment, $BC_{el}$ systematically under-estimates $\theta$. Figure 5 shows the typical discrepancy between both results: ABC and $BC_{el}$ agree on the mutation rate $\theta$, but the $BC_{el}$ estimation of $\tau$ is more accurate, see also Table 1.

In the second experiment, we consider three populations, see Figure 4 *(right)*: the last two populations diverged at time $\tau_1$ and their common ancestral population diverged from the first population at time $\tau_2$. The sample comprises thirty diploid individuals per population genotyped at a hundred independent loci. In contrast to the first experiment, all components of the composite scores are computed here by summing over all pairs of genes whatever the population to which they belong. The results given in Table 1 show that ABC and $BC_{el}$ mainly agree on both parameters $\theta$ and $\tau_1$, but $BC_{el}$ is slightly more accurate than ABC on $\tau_2$.

Table S1 gives a comparison of the computing times for both algorithms, showing the difference of magnitudes between them. This is due to the simulation of the the simulated datasets for ABC: While this difference should not be over-interpreted, it signals a potential for self-assessment and testing that is missing for ABC methods.

Table 1: Comparison of the original ABC and BC$_{el}$ on 100 Monte Carlo replicates. We use two point estimates of the parameters: **(1)** posterior mean and **(2)** posterior median, and measured the error between the estimation and the "true" value used to generate the observation with **(1)** the root mean square error in the case of the posterior mean and **(2)** the median absolute deviation in the case of the posterior median. We also compare credible intervals (with probability 0.8) through the proportion of Monte Carlo replicates in which the "true" value falls into this interval.

| | **First experiment** | | | | | |
|---|---|---|---|---|---|---|
| | Root Mean Square Error of posterior mean | | Median Absolute Deviation of posterior median | | Coverage of the credible interval with probability 0.8 | |
| | ABC | BC$_{el}$ | ABC | BC$_{el}$ | ABC | BC$_{el}$ |
| $\theta$ | 0.0971 | 0.0949 | 0.071 | 0.059 | 0.68 | 0.81 |
| $\tau$ | 0.315 | 0.117 | 0.272 | 0.077 | 1.0 | 0.80 |
| | **Second experiment** | | | | | |
| | Root Mean Square Error of posterior mean | | Median Absolute Deviation of posterior median | | Coverage of the credibility interval of probability 0.8 | |
| | ABC | BC$_{el}$ | ABC | BC$_{el}$ | ABC | BC$_{el}$ |
| $\theta$ | 0.0593 | 0.0794 | 0.0484 | 0.0528 | 0.79 | 0.76 |
| $\tau_1$ | 0.472 | 0.483 | 0.320 | 0.280 | 0.88 | 0.76 |
| $\tau_2$ | 29.6 | 4.76 | 4.13 | 3.36 | 0.89 | 0.79 |

## 4 Discussion

When compared with ABC methods, the (often) significant time savings provided by BC$_{el}$ due to the lack of pseudo-sample simulation may open wider ranges for processing models involving complex likelihoods. For instance, in population genetics, ABC is severely hindered by the time spent simulating a dataset when modelling isolation by distance in a continuously distributed population, or when studying a large set of SNP markers even on quite simple evolution scenarios. Moreover, when the dataset is composed of large sets of markers, the summary statistics proposed in ABC (in DIY-ABC, these are averages of some quantitative statistics over all loci) ignores some (statistical) information, while BC$_{el}$ manages to recover most of it, more specifically to estimate divergence on large datasets. Improvements in accuracy of estimation and computational efficiency are also possible in other contexts as illustrated in the range of examples given above.

Even when BC$_{el}$ requires the same computing time as ABC, it uses the outcome in a very different perspective and provides a benchmark likelihood that helps in evaluating the pertinence of the ABC approximation, as illustrated in the gamma point process of SI.

We acknowledge that a caveat of the empirical likelihood is that it requires a careful choice of the constraint (1). Those pivotal quantities have to be connected to the parameter in an identifying way, which may require complex manipulations as in the gamma process case or even be impossible. However, repeated experimentation is often available, as illustrated by the normal example and the population genetic experiments (where we computed the composite score on both a restricted set of pairs and all pairs of genes). Checking for the accuracy of the approximation means that a constraint in BC$_{el}$ should be tested on simulated datasets in controlled experiments where the true parameters are known, although much less than in ABC runs. Then we can test coverage of credibility intervals, and measure the error of various point estimates based on the output of the scheme.

# Acknowledgments

The last two authors wish to thank Jean-Marie Cornuet for his help and availability. Their work has been partly supported by the Agence Nationale de la Recherche (ANR) through the 2009–2012 project Emile. The third author is grateful to Patrice Bertail, Chris Drovandi, Brunero Liseo, and Art Owen for useful discussions. Comments and suggestions from the whole PNAS editorial board greatly contributed to improve both the presentation and the scope of the paper.

# Supplementary information (SI)

### Convergence of the empirical likelihood approximation

The validation of the empirical likelihood approximation is provided by Theorem 3.4 of (1), which establishes an extension of Wilk's theorem to the EL likelihood ratio. (Note that $n^{-n}$ is the maximum of $L_{\text{el}}$.)

**Theorem** *Let $X, Y_1, \ldots, Y_n$ be independent random vectors with common distribution $F_0$. For $\theta \in \Theta$, let $h(X, \theta) \in \mathbb{R}^s$. Let $\theta_0 \in \Theta$ be such that $\text{Var}(h(Y_i, \theta_0))$ is finite and has rank $q > 0$. If $\theta_0$ satisfies $\mathbb{E}(h(X, \theta_0)) = 0$, then $-2\log\left(\dfrac{L_{\text{el}}(\theta_0|Y_1, \ldots, Y_n)}{n^{-n}}\right) \to \chi^2_{(q)}$ in distribution when $n \to \infty$.*

We also reproduce here an illuminating comment from Art Owen: *"The interesting thing about Theorem 3.4 is what is not there. It includes no conditions to make $\hat\theta$ a good estimate of $\theta_0$, nor even conditions to ensure a unique value for $\theta_0$, nor even that any solution $\theta_0$ exists. Theorem 3.4 applies in the just determined, over-determined, and under-determined cases. When we can prove that our estimating equations uniquely define $\theta_0$, and provide a consistent estimator $\hat\theta$ of it, then confidence regions and tests follow almost automatically through Theorem 3.4."*.

### Pairwise composite likelihoods in population genetics

We detail here the derivation of the composite likelihoods used for the version of the $\text{BC}_{\text{el}}$ algorithm implemented in the case of the population genetics study.

**Two genes from the same deme** First, we recall that we scale the time axis so that a pair of genes of the same deme coalesces at a random time with an exponential distribution with rate 1. We now consider a given locus and two microsatellite genes from our sample that come from the same deme. We denote their respective allelic state by $x_1$ and $x_2$. Their most recent common ancestor (MRCA) dates back to a time $T$, where $T \sim \mathcal{E}(1)$. We assume that the mutation rate, namely $\theta/2$ does not vary along the whole history of our populations. Therefore, conditioned on $T$, the number of mutations between $x_i$ ($i = 1, 2$) and the MRCA is distributed according to a Poisson distribution with mean $\theta T/2$. Hence, conditional on $T$, the number $N_0$ of mutations between $x_1$ and $x_2$ is a Poisson variable with mean $\theta T$ and

$$\mathbb{E}\left[u^{N_0}\big|T\right] = \exp\{\theta T(u-1)\}.$$

Thus

$$\begin{aligned}\mathbb{E}\left[u^{N_0}\right] &= \int_0^\infty e^{-t} \exp\{\theta t(u-1)\}\, dt \\ &= \frac{1}{1+\theta(1-u)} = \frac{1/(1+\theta)}{1-\theta/(1+\theta)u}.\end{aligned}$$

i.e., $N_0 \sim \mathcal{G}e(\theta/(1+\theta))$, the geometric distribution with positive weight at 0. Finally, the difference between both genes is the accumulation over the $N_0$ mutations, i.e.,

$$x_2 - x_1 = \sum_{k=1}^{N_0} \epsilon_k$$

where the $\epsilon_k$'s are iid Rademachers ($\pm 1$ with equal probability). Thus,

$$\mathbb{E}[e^{i\zeta(x_2-x_1)}] = \mathbb{E}\left\{\mathbb{E}\left[e^{i\zeta(x_2-x_1)}\big|N_0\right]\right\} = \mathbb{E}\left\{\prod_{k=1}^{N_0}\mathbb{E}[\exp(i\zeta\epsilon_k)]\right\}$$

$$= \mathbb{E}\left[(\cos\zeta)^{N_0}\right] = \sum_{n=0}^{\infty}(1-p)(p\cos\zeta)^n$$

since $N_0 \sim \mathcal{G}e(p)$ with $p = \dfrac{\theta}{1+\theta}$

$$= \frac{1-p}{1-p\cos\zeta} = \frac{1}{1+\theta(1-\cos\zeta)} \tag{4}$$

which proves that the pairwise likelihood is

$$\ell_2(x_1, x_2|\phi) = \frac{1}{\sqrt{1+2\theta}}\rho(\theta)^{|x_2-x_1|}, \tag{5}$$

with $\rho(\theta) = \theta/(1+\theta+\sqrt{1+2\theta})$.

**Two genes from different demes** We now consider two genes that come from two different demes that diverged from an ancestral deme at time $\tau$ in the past. We denote the allelic state of the two ancestors at time $\tau$ by $x_1^0$ and $x_2^0$, respectively. Then, $x_1 - x_2 = (x_1 - x_1^0) + (x_1^0 - x_2^0) + (x_2^0 - x_2)$, where $x_1^0 - x_2^0$ follows a distribution whose Fourier transform is given by (4), while $(x_j - x_j^0)$, $j=1,2$ are iid., whose distribution is given by the difference of two allelic states separated by a fixed time $\tau$. This distribution is derived in Equation (3) of (2):

$$\mathbb{P}(x_j - x_j^0 = \delta) = e^{-\tau\theta/2}I_\delta(\tau\theta/2),$$

where $I_\delta$ denotes the $\delta$th-order modified Bessel function of the first kind, given by ($n \geq 0$)

$$I_{-n}(z) = I_n(z) = \sum_{k=0}^{\infty}\frac{(z/2)^{n+2k}}{k!(n+k)!}.$$

Using the independence between $(x_1 - x_1^0)$ and $(x_1 - x_1^0)$, we obtain

$$\mathbb{P}\big((x_1 - x_1^0) + (x_1 - x_1^0) = \delta\big) = e^{-\tau\theta}I_\delta(\tau\theta). \tag{6}$$

We then retrieve this distribution by computing Fourier transforms in the same vein as above. First, we note that the number $N_1$ of mutations between $x_1$ and its ancestor at time $\tau$ is a Poisson variate with parameter $\tau\theta/2$. And,

$$\mathbb{E}[e^{i\zeta(x_1^0-x_1)}] = \mathbb{E}\left\{\mathbb{E}\left[e^{i\zeta(x_1^0-x_1)}\big|N_1\right]\right\} = \mathbb{E}\left(\prod_{k=1}^{N_1}\mathbb{E}(\exp(i\zeta\epsilon_k))\right)$$

$$= \mathbb{E}\left[(\cos\zeta)^{N_1}\right] = \sum_{n=0}^{\infty}e^{-\tau\theta/2}\frac{(\tau\theta/2)^n}{n!}(\cos\zeta)^n$$

since $N_1 \sim \mathcal{P}o(\tau\theta/2)$

$$= \exp\{\tau\theta(\cos\zeta - 1)/2\}. \tag{7}$$

Finally, the distribution of $x_2 - x_1$ is a (discrete) convolution product between the distributions given by (5) and (6) which yields

$$\ell_2(x_1, x_2|\phi) = \frac{e^{-\tau\theta}}{\sqrt{1+2\theta}}\sum_{k=-\infty}^{+\infty}\rho(\theta)^{|k|}I_{|x_1-x_2|-k}(\tau\theta).$$

## Quantile estimation

Examples of quantile distributions are the three-, four- and five-parameter Tukey's lambda distributions and their generalizations and the Burr family of distributions; particular examples include the $g$-and-$h$ and $g$-and-$k$ distributions (3, 4, 5, 7).

Proposed methods for estimation of quantile distributions include maximum likelihood estimation using numerical approximations to the likelihood (7, 8, 9), moment matching (10, 11), location and scale-free shape functionals (12), percentile matching (6), quantile matching (13) and, more recently, ABC (14, 15, 16). Sequential Monte Carlo approaches for multivariate extensions of the $g$-and-$k$ have also been proposed (37).

There has been a number of ABC approaches proposed for this problem. For example, (15) adopted the ABC-MCMC algorithm of (18), in which draws of $\theta$ are based on a Metropolis algorithm with a Gaussian proposal distribution, and are accepted based on the rule $\rho(S(D), S(D\prime)) < \epsilon)$, where $D$ is the entire set of order statistics, $\rho$ is the Euclidean norm and $\epsilon$ is heuristically chosen after inspection of a histogram of $\rho(S, S\prime)$ obtained from a preliminary run using a very large value of $\epsilon$. This approach has recently been improved by (16) through more sophisticated MCMC approaches, the use of regression summary statistics for $D$ based on percentiles and their powers, and more automated choices of $\epsilon$. However, they still maintain a form of distance-based measure $\rho(S, S\prime)$ in accepting $\theta$.

## Normal estimation

Figures S1–S3 evaluate the impact on the posterior distribution approximation of increasing the number of constraints in the empirical likelihood definition. Since this is a formal example, the true posterior distribution is available.

## Quantile estimation

The $BC_{el}$ experiment involved evaluation of Algorithm 1 for estimation of the parameters of the $g$-and-$k$ distribution using the two values of $\theta = (A, B, g, k)$, namely $\theta_N = (0, 1, 0, 0)$, which corresponds to the standard normal distribution, and $\theta_A = (3, 2, 1, 0.5)$; which was chosen by (36) as 'an interesting, far-from-normal distribution. The simulation experiment comprised multiple repetitions of $BC_{el}$ using different combinations of sample size, $n = (100, 500)$, number of iterations, $M = (1000, 5000, 10000)$, and number of constraints ($p = 3, 4, 4, 5, 9$), corresponding to percentile sets $(0.2, 0.5, 0.8)$, $(0.2, 0.4, 0.6, 0.8)$, $(0.1, 0.4, 0.6, 0.9)$, $(0.1, 0.25, 0.5, 0.75, 0.9)$ and $(0.1, 0.2, ..., 0.9)$. Two sets of priors were considered for $(A, B, g, k)$: $U[0, 5]^4$ (denoted as $P_1$) and $U(-5, 5).U(0, 5), U(-5, 5), (-0.1, 1)$ (denoted as $P_2$). Although the priors were set independently for each element of $\theta$, the four elements were drawn together at each iteration of the algorithm, so that the same importance weight $\omega_i$ was attached to the values $A^{(i)}, B^{(i)}, g^{(i)}, k^{(i)}$ drawn in the $i$th iteration. The experiment was replicated ten times with different draws of samples of size $n$. Posterior means and standard deviations were computed for each parameter, and the overall goodness of fit to the true curve was assessed by comparing the true quantiles at $(0.05, 0.10, ..., 0.95)$ with two measures: the estimated mean at each quantile (denoted by RSSm) and the average of the estimated quantile for each importance sample (RSSt).

Boxplots of the posterior means and standard deviations are shown in Figure S4 and Figure S5, respectively, for the four parameters, based on $\theta_A$, prior $P_2$ and the 20 replicates for $M = (5000, 10000)$, for ten of the trials: $p = 3, 4, 4, 5, 9$ for $n = 100$ (trials 1-5) and $n = 500$ (trials 6-10). Boxplots for the corresponding overall goodness of fit measures (RSSm, RSSt) are given in Figure S6.

## Superposition of point processes

(41) discuss an alternative to ABC, using fractional design and linear interpolation. While their purpose is the non-Bayesian processing of models with intractable likelihood functions, they propose as their main example a model consisting in the superposition of $N$ renewal processes with waiting times $\tau_{ij}$ ($i = 1, \ldots, M$, $j = 1, \ldots$) distributed as $\mathcal{G}(\alpha, \beta)$ variables, when $N$ is unknown. The renewal processes are thus

$$\zeta_{i1} = \tau_{i1}, \ \zeta_{i2} = \zeta_{i1} + \tau_{i2}, \ \ldots$$

and the observations are made of the first $n$ values of the $\zeta_{ij}$'s,

$$z_1 = \min\{\zeta_{ij}\}, \ z_2 = \min\{\zeta_{ij}; \zeta_{ij} > z_1\}, \ \ldots,$$

ending with

$$z_n = \min\{\zeta_{ij}; \zeta_{ij} > z_{n-1}\}.$$

This model offers an interesting testing ground for $BC_{el}$ in that the data points $z_t$ are neither iid nor Markov. It is however possible to recover and exploit an iid structure in this case by first simulating a pseudo-dataset, $(z_1^\star, \ldots, z_n^\star)$, as in ABC settings, and then deriving a sequence of renewal processes indicators $(\nu_1, \ldots, \nu_n)$, as

$$z_1^\star = \zeta_{\nu_1 1}, \ z_2^\star = \zeta_{\nu_2 j_2}, \ \ldots, z_n^\star = \zeta_{\nu_n j_n}.$$

These indicators are thus distributed from the prior distribution on the $\nu_t$'s and an iid sample of $\mathcal{G}(\alpha, \beta)$ variables can be derived from those indicators and the genuine data, leading to an associated empirical likelihood. As shown on Figure S7, when applied to a simulated dataset (as in (41)), the empirical likelihood approximation produces a better approximation than the corresponding ABC solution based on the same statistics as (41) (for exactly the same computational cost).

### Time gains in population genetic models

In general, the speed of executing an ABC algorithm depends on many factors, including:

- one's ability to program an efficient simulator from the model distribution and to compute the selected summary statistics,
- the choice of the threshold $\epsilon$
- and the size of the Monte Carlo sample, denoted $M$,

and the speed of $BC_{el}$ depends on:

- the difficulty to optimize under the constraints [1] (in the population genetics examples, this is not straightforward because of the Bessel functions and of the various series involved when the two individuals are not in the same deme),
- and the size $M$ of the Monte Carlo sample.

While producing many simulations from the model distribution often is the stumbling block for ABC algorithms, the selection of the constraints [1] and the time requirements of the optimization step are both highly variable and delicate to quantify. While all the experiments described in this paper induced no inflation in computing time and mostly significant reductions, we cannot exclude the possibility of $BC_{el}$ requiring more computing time than ABC.

Both population genetic experiments conducted in this paper analyse datasets with a large number of loci (one hundred). Thus ABC, which requires simulations of all loci to produce a simulated dataset, is quite time consuming and particularly so when the evolutionary scenario is more complex than the one in the first experiment. We compare here the computing times required by our implementation of the $BC_{el}$-AMIS sampler and by DIYABC (3) on an Intel Xeon W3680 Plateform with GNU/Linux. Both methods were parallelized over five among the six cores of this CPU with the OpenMP API. Table 2 exhibits computation time averages on ten replicates of the estimation.

Table 2: Computing times for DIYABC and $BC_{el}$ in both population genetics experiments

| Experiment | ABC (DIYABC software) | $BC_{el}$ ($BC_{el}$-AMIS code) |
|---|---|---|
| 1 | 21 min | 24 sec |
| 2 | 16 hours | 55 sec |

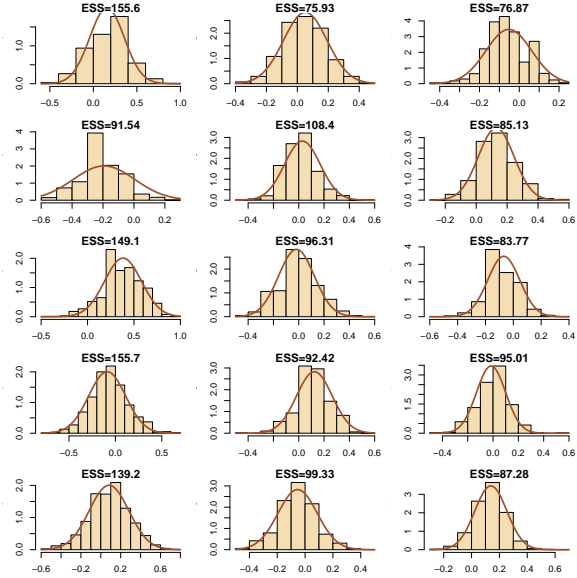

Figure 6: Comparison of the true posterior on the normal mean (solid lines) with the empirical distribution of weighted simulations resulting from Algorithm $BC_{el}$. The normal sample sizes are 25 (column 1), 50 (column 2) and 75 (column 3), the number of simulated $\theta$'s is $10^3$ and the effective samples sizes $M^{ESS}$ are given on top of each histogram. The constraint is on the first moment of the dataset.

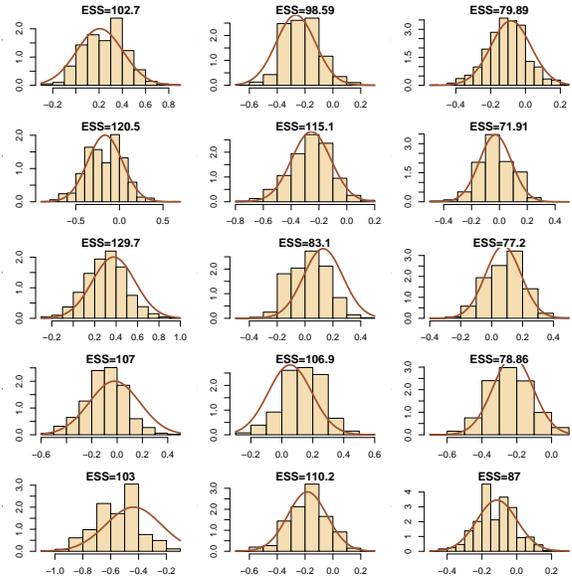

Figure 7: Comparison of the true posterior on the normal mean (solid lines) with the empirical distribution of weighted simulations resulting from Algorithm $BC_{el}$. The normal sample sizes are 25 (column 1), 50 (column 2) and 75 (column 3), the number of simulated $\theta$'s is $10^3$ and the effective samples sizes $M^{ESS}$ are given on top of each histogram. The constraint is on the first two moments of the dataset.

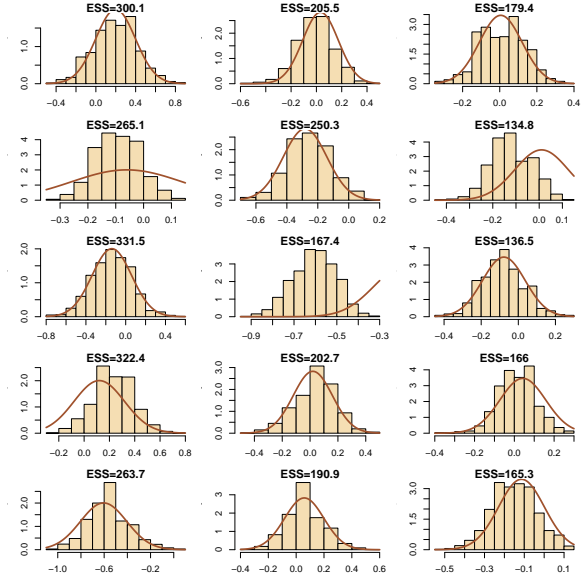

Figure 8: Comparison of the true posterior on the normal mean (solid lines) with the empirical distribution of weighted simulations resulting from Algorithm $BC_{el}$. The normal sample sizes are 25 (column 1), 50 (column 2) and 75 (column 3), the number of simulated $\theta$'s is $10^3$ and the effective samples sizes $M^{ESS}$ are given on top of each histogram. The constraint is on the first three moments of the dataset.

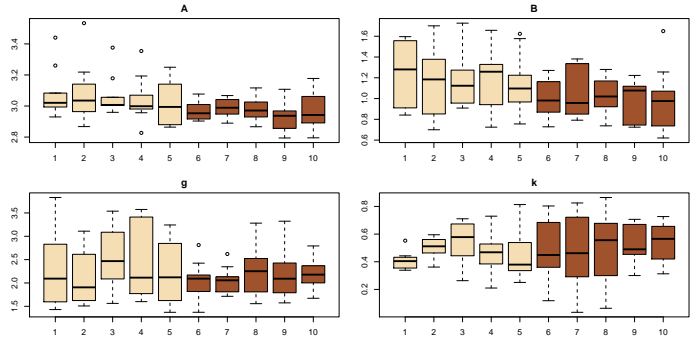

Figure 9: Boxplots of the variations of the posterior means of the four parameters of the $g$-and-$k$ distribution, based on $BC_{el}$ approximations, for $n = 100$ observations (1 to 5) and $n = 500$ observations (6 to 10), based on $M = 10^4$ simulations and 10 replications. The moment conditions used in the $BC_{el}$ algorithm are quantiles of order $(0.2, 0.5, 0.8)$ (1 and 6), $(0.2, 0.4, 0.6, 0.8)$ (2 and 7), $(0.1, 0.4, 0.6, 0.9)$ (3 and 8), $(0.1, 0.25, 0.5, 0.75, 0.9)$ (4 and 9), and $(0.1, 0.2, \ldots .0.9)$ (5 and 10).

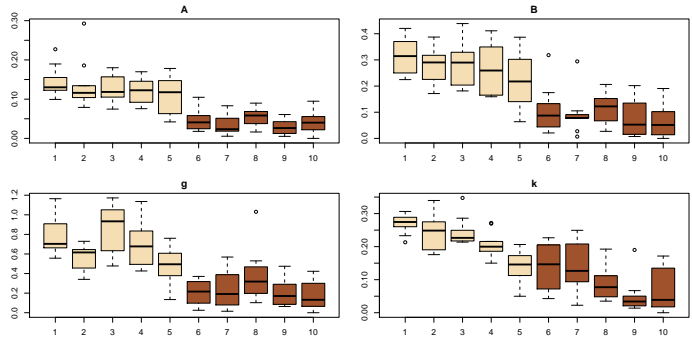

Figure 10: Same graph as Figure 9 for the standard deviations.

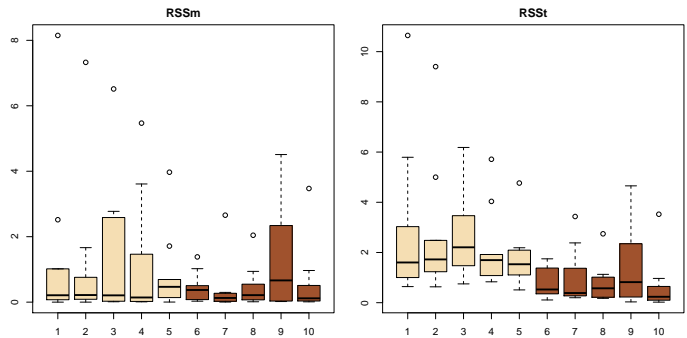

Figure 11: Goodness of fit measures (RSSm, RSSt) for the same experiment as in Figure 9.

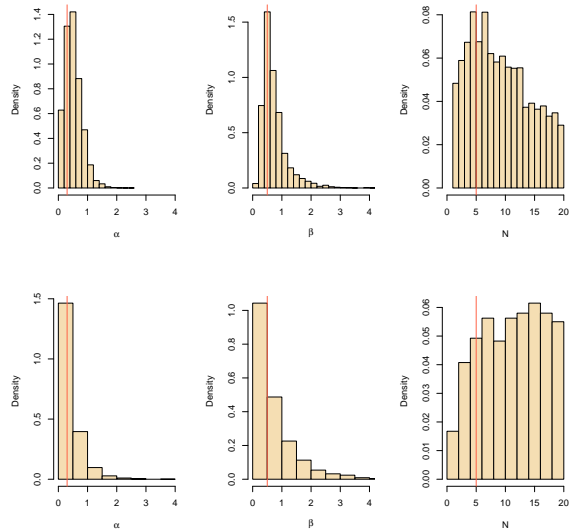

Figure 12: Approximate posterior distributions of the parameters $(\alpha, \beta, N)$ for the superposition gamma process model, using a simulated dataset of 90 observations, with $\alpha = .5, \beta = .8$, and $N = 5$: *(top)* $BC_{el}$ output; *(bottom)* ABC output.

21